 \documentclass[preprint2]{aastex}

\shorttitle{Stellar Variability in the Globular Cluster Terzan 5}
\shortauthors{Edmonds et al.}

\begin{document}

\title{Stellar Variability in the Metal-Rich, Obscured Globular Cluster
Terzan 5 \footnote{Based on observations with the NASA/ESA Hubble Space
Telescope obtained at the Space Telescope Science Institute, which is
operated by Association of Universities for Research in Astronomy,
Incorporated, under NASA contract NAS 5-26555}}

\author{Peter D. Edmonds \& Jonathan E. Grindlay}
\affil{Harvard College Observatory, 60 Garden Street, Cambridge, MA 02138}
\email{pedmonds@cfa.harvard.edu,josh@cfa.harvard.edu}

\author{Haldan Cohn \& Phyllis Lugger}
\affil{Department of Astronomy, Indiana University, Swain West 319\\
Bloomington, IN 47405}
\email{cohn@indiana.edu,lugger@indiana.edu}

\begin{abstract}

We present the results of a search for variability in and near the core of
the metal-rich, obscured globular cluster Terzan 5, using NICMOS on {\it
HST}. This extreme cluster has approximately solar metallicity and a
central density that places it in the upper few percent of all clusters.
It is estimated to have the highest interaction rate of any galactic
globular cluster. The large extinction towards Terzan 5 and the severe
stellar crowding near the cluster center present a substantial
observational challenge. Using time series analysis we discovered two
variable stars in this cluster. The first is a RRab Lyrae variable with a
period of $\sim$0.61 days, a longer period than that of field stars with
similar high metallicities. This period is, however, shorter than the
average periods of RR Lyraes found in the metal-rich globular clusters NGC
6441, NGC 6388 and 47 Tuc. The second variable is a blue star with a
7--hour period sinusoidal variation and a likely orbital period of 14
hours. This star is probably an eclipsing blue straggler, or (less likely)
the infrared counterpart to the low mass X-ray binary known in Terzan
5. Due to the extreme crowding and overlapping Airy profile of the IR PSF,
we fall short of our original goal of detecting CVs via P$\alpha$ emission
and detecting variable infrared emission from the location of the binary
MSP in Terzan 5.\\

\end{abstract}



\section{Introduction}

The study of variable stars in globular clusters has a long, rich history.
Studies of RR Lyrae variables, long period variables and Population II
cepheids have addressed fundamental questions about stellar evolution,
globular cluster evolution and the distance scale. More recently,
observations of variable stars such as millisecond pulsars (MSPs),
cataclysmic variables (CVs) and eclipsing binaries have been motivated by
studies of stellar interactions and globular cluster dynamics. For example,
during stellar encounters binaries are expected to become more tightly
bound and give the energy difference up to passing stars, a process that
can support a cluster against core collapse.

Observations with {\it HST} have been crucial in overcoming the significant
crowding present in such dense stellar environments.  For example, {\it
HST} observations of the center of the massive globular cluster 47 Tucanae
have discovered two likely CVs (Paresce and de Marchi 1994 and Shara et
al. 1996), nine eclipsing binaries (Edmonds et al. 1996) and 6 pulsating
blue stragglers (BSs) (Gilliland et al. 1998). Observations of the
collapsed--core cluster NGC 6397 have discovered 5 CVs (Cool et al. 1995,
Grindlay et al. 1995; Edmonds et al. 1999; Grindlay 1999) and 3 likely
low-mass WDs (Cool et al. 1998 and Edmonds et al. 1999) that may also be
MSPs.

{\it HST} has also been used to obtain the first high quality images and
color magnitude diagrams (CMDs) of NGC 6441 and NGC 6388, two partially
obscured clusters with surprisingly rich blue horizontal branches (HBs)
when their high metallicity is taken into account (Rich et al. 1997). Most
metal--rich clusters have red, stubby HBs, in agreement with standard
theories of stellar evolution. As metallicity decreases HBs usually become
increasingly blue, but some clusters, such as NGC 6441 and NGC 6388,
contradict this pattern and a second parameter is assumed to affect HB
morphology.  The cause of this ``2nd parameter effect'' is a subject of
vigorous debate, but some intriguing clues have appeared. Both NGC 6441 and
NGC 6388 appear to have HBs that are brighter than the HBs in metal--poor
clusters. Variability searches in these clusters have subsequently shown
that the RR Lyrae stars have significantly longer periods than expected for
metal-rich stars (Layden et al. 1999 and Pritzl et al. 2000), as does the
RR Lyrae V9 in 47 Tuc (Carney et al. 1993). This apparently new subclass of
long period, metal--rich RR Lyraes may show the lack of universality of the
luminosity--metallicity relation for RR Lyraes and may require the
establishment of a third Oosterhoff group (Pritzl et al. 2000).  

The globular cluster Terzan 5, the subject of this paper, represents an
excellent laboratory for the studies introduced above. Because its
metallicity is close to solar (Ortolani, Barbuy and Bica 1996; Cohn et
al. 2000) it is ideal for studies of HB morphology and RR Lyrae behavior in
a metal--rich globular cluster. It also has the highest predicted rate of
stellar collisions of any galactic globular cluster (Verbunt \& Hut 1987)
making it a laboratory for stellar dynamics. It is already known to contain
a LMXB (Johnston, Verbunt and Hasinger 1995), a binary MSP (Lyne et
al. 1990; Nice \& Thorsett 1992), one other pulsar and a large number of
unresolved pulsars (Fruchter \& Goss 2000). Unfortunately Terzan 5 is
highly reddened, with E(B-V) = 2.49 (Ortolani et al. 1996) and because of
its large mass the crowding near cluster center, where the binaries are
expected to be concentrated, is a formidable problem.

Here, we describe an attempt to overcome the high reddening and crowding in
Terzan 5 by observing it in the IR with NICMOS on {\it HST}. Our aims were
(1) to search the center of the cluster for CVs and the IR counterpart of
the LMXB using wide (F187W) and narrow-band (F187N) NICMOS filters centered
on Paschen-$\alpha$ (P$\alpha$), and (2) to search for variability from the
binary MSP. The general photometric results are described in Cohn et
al. (2000). In this paper we describe a detailed search for variability in
two fields, one covering the center of the cluster and another containing
the binary MSP. Although we fall short of detecting CVs because of crowding
and no evidence is found for variability of any of the objects lying in the
error circle of the MSP, we do detect an interesting RR Lyrae variable and
a second variable that is likely to be an eclipsing BS, or (less likely)
the IR counterpart of the LMXB.

\section{Observations and Analysis}

\subsection{Data Reduction}

The Terzan 5 NICMOS observations described here were obtained on June 2nd
and 3rd, 1998.  Two pointings of 4 orbits each were obtained. Pointing (1)
directed the NIC2 camera ($19.2''\times 19.2''$ field of view) at the
center of the cluster and pointing (2) directed the NIC1 camera
($11.5''\times 11.5''$ FOV) at the eclipsing MSP. Pointing (1) was adopted
for $\sim$75\% of each orbit and the telescope then slewed to pointing (2) for
the remainder of each orbit. All three NICMOS cameras were used in parallel
for these observations, but only the NIC2 data for pointing (1) and the
NIC1 data for pointing (2) are of interest and are described here.

Within pointing (1) a dither pattern was adopted, with different pointing
positions for each of the 8 orbits. The pattern consisted of 8 out of 9
points of a 6 pixel by 6 pixel square dither, with 3 pixel
steps. Superimposed on this box pattern were sub-pixel offsets in 1/8 pixel
steps.  Both of these factors were designed to improve the quality of the
photometry, at the expense of the quality of the time series. The box
pattern corrects the photometry for image defects such as bad pixels and
flat field errors and the sub-pixel offsets improve the effective sampling
of the images, for better resolution. For the MSP pointing, no dithering
was used to maximize the time series quality, but a different field
position was used for the second set of 4 orbits to guard against
accidental positioning on a bad pixel.

For the central NIC2 observations, F110W, F187W and F187N were used for
each of the 8 orbits. The F187W and F187N filters are the infrared
analogues of $R$ and H$\alpha$ respectively and the F110W filter is similar
to $J$. The latter was included to give color information for the observed
stars.  For the NIC1/MSP field only F140W images were obtained. This filter
was chosen because it gave the highest S/N and the greatest chance of
detecting variability from the MSP, in particular evidence of orbital
variability from the secondary, as in the system PSR1957+20 (Djorgovski \&
Evans 1988; Callanan, van Paradijs \& Rengelink 1995). For the remainder
of this section we will discuss only the central NIC2 observations.

The NICMOS data were obtained in MULTIACCUM mode, which provides
intermediate, non-destructive readouts during a single observation. For the
F110W and F187W filters the `STEP32' sequence was used, with 17 or 18
samples. The readout times were 0.3s, 0.6s, 1.0s, 2.0s, 4.0s, 8.0s, 16.0s,
32.0s, 64.0s, etc (the remaining times in 32.0s steps). For the F187N
filter the `STEP128' sequence was used with the same readout times as above
except extending up to 128s readouts (with 22 samples). Standard processing
was applied via the STScI pipeline. Here, a problem was discovered with
the processed F187N images, with a large number of stars found to contain
pixels with unusually small signal levels.  With guidance from Mark
Dickinson, Torsten Boecker and Howard Bushouse at STScI, it was discovered
that the saturation thresholds for the F187N processing were set
incorrectly. Lowering the saturation thresholds by 5-10\% and reprocessing
using CALNICA repaired almost 100\% of these faulty pixels.

Customized software was used to convert the accumulated data into
individual exposures suitable for time series analysis. The data quality
files were used to remove CRs and to avoid using saturated pixels.  After
blinking the individual images, pixels affected by `grot' (probably small
flecks of paint) were flagged as bad pixels and the bad pixel list was
updated. Single bad pixels were corrected or `cleaned' by adopting the
average of the 4 adjacent pixels, but no attempt was made to correct groups
of two or more adjacent bad pixels.

\subsection{Time Series Analysis}

Our star list for differential photometry was based on the work of Cohn et
al. (2000) who averaged and combined the 8 separate images for each filter
into single, over-sampled images using the STSDAS drizzle routine.  The
F110W filter, which gave the deepest, best resolved final image was used to
calculate the master star list used here (containing 3091 stars).

Two standard photometry techniques were used to produce time series:

(1) Aperture photometry: A plate solution determined from a sample of
bright stars was applied at each time, to calculate the coordinate shifts
to be applied to the master coordinate list. Aperture photometry was then
carried out using DAOPHOT under IRAF, with an aperture radius of 2.5
pixels, roughly matching the radius of the first Airy minima for the F187
data. To improve the signal-to-noise (S/N) ratio of the intensity
measurements and minimize the contribution from near neighbors, the signal
in each pixel in each aperture was weighted by a simple model of the PSF.
Sky values were determined from the mode of the counts in an annulus around
each star.

(2) Point spread function (PSF) fitting: The extreme crowding of the
central Terzan 5 field, especially the large number of bright giants, the
complicated nature and large radial extent of the PSF and the marginal
sampling of the individual images caused problems with PSF-fitting for the
faintest stars. In particular, many faint objects in the master star list
were not detected in the F110W images (the most undersampled data) using
ALLSTAR.  Therefore, we experimented with a smaller sized list of 1702
stars containing objects found in PSF--fitting of both the F110W and F187W
images (Cohn et al. 2000).

We examined the counts for each star as a function of time for the longest
readout times (32s or 128s), and found a high correlation between the time
series of different stars, with a linear falloff in the counts as a
function of time during an exposure, caused, we believe, by `persistence'
(Skinner et al. 1998).  To remove this artifact we calculated the average
time series for bright non-saturated stars and normalized each time series
by this average function.  Zero-points based on the photometry of Cohn et
al. (2000) were used.

We then compared the two photometry techniques using the root-mean-square
(rms) values calculated for each star. Only the exposures with the longest
sample times (either 32s or 128s) were included in this calculation.
Aperture photometry gave significantly better results than PSF fitting for
the F110W data and comparable or slightly better results for the F187W and
F187N data.  Since the aperture photometry results were superior we will
quote them unless otherwise noted.

\begin{figure}[hbt]
\vspace*{2cm}
\plotone{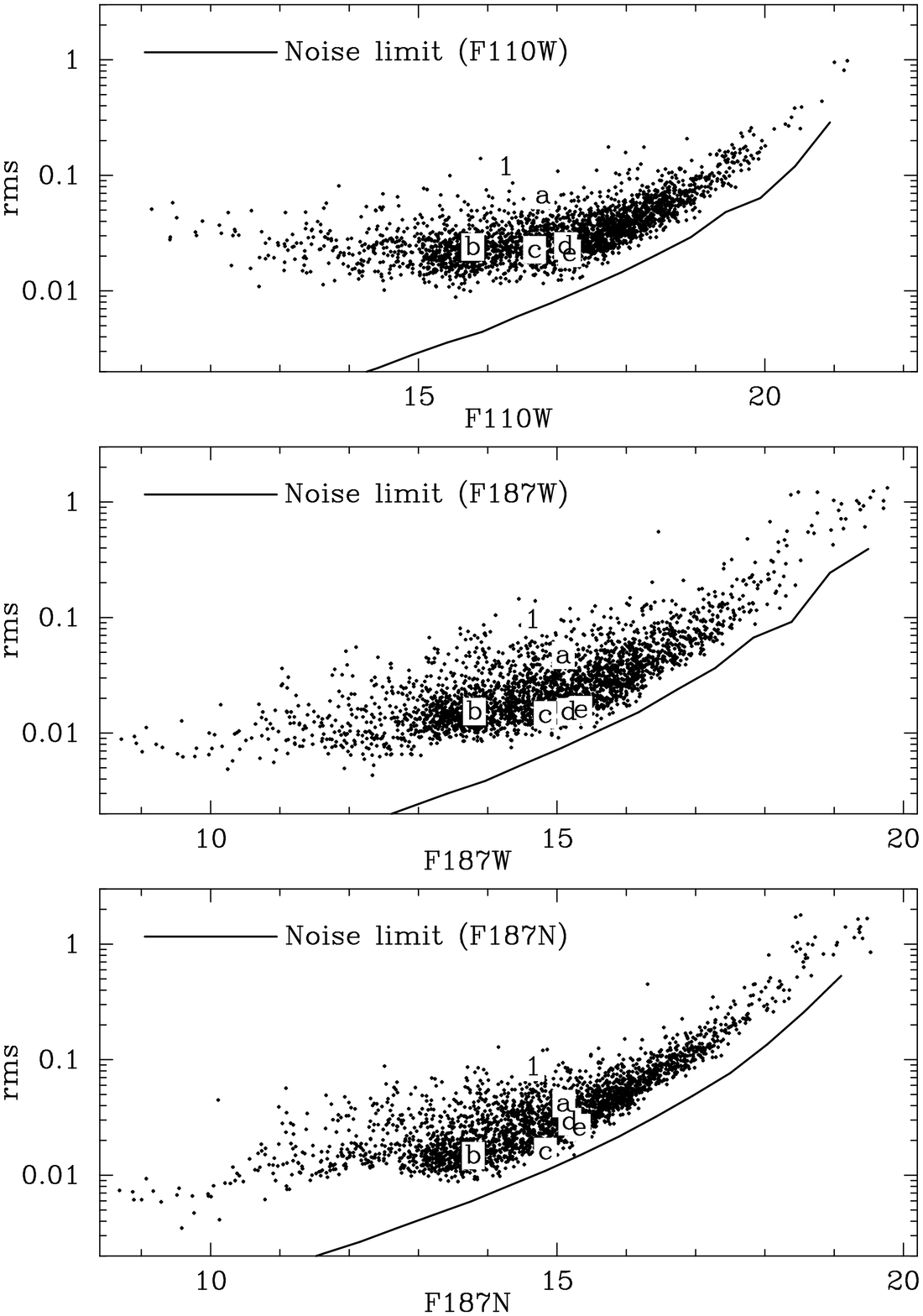}
\caption{Plots of rms versus magnitude for the three different NIC2 filters
used. The solid line shows the Poisson limits for aperture photometry. The
symbols represent the variable V1 (`1') and its near neighbors (`a'-`e').}
\label{fig.allmagrms}
\end{figure}

Figure \ref{fig.allmagrms} shows plots of rms versus F110W, F187W and F187N
magnitudes.  The solid line shows the theoretical rms based on the limits
from photon noise and readout noise (Kjeldsen \& Frandsen 1992).  We have
plotted the average rms limit in 0.5 mag wide bins in each case.  At faint
magnitudes our measured rms values are only factors of 2--3 above the
Poisson limit, a reasonable result given the large amount of dithering
adopted, and other systematic effects.

\begin{figure}[hbt]
\vspace*{-5cm}
\plotone{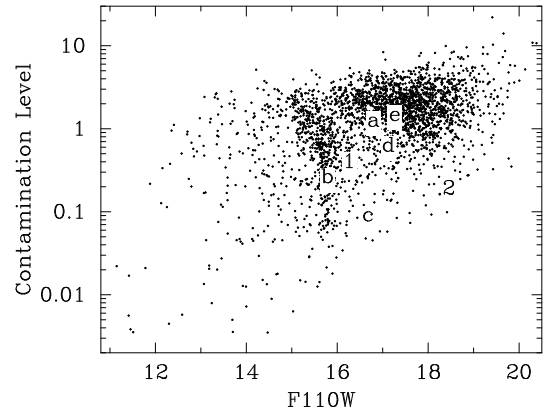}
\vspace*{-1.5cm}
\caption{Contamination level plotted against F110W. The same symbols as
Figure \ref{fig.allmagrms} are used except we also include the variable
V2 (`2').}
\label{fig.contam}
\end{figure}

Because the images are crowded we have examined the degree of overlap
between stars. We define the `contamination' level (CL) for any star to be
the amount of light from other stars falling within its aperture (for
example, CL = 0 implies a star is completely isolated and CL = 1 implies
equal contribution from the star and neighbors).  The CLs for the F110W
filter are shown in Figure \ref{fig.contam}. The median CL is 1.70, showing
the high level of crowding in the data.

To search for variables we concentrated on time series analysis.  We placed
the stars in order of decreasing brightness, divided the stars into bins
containing 50 stars and searched for stars with unusually large rms values
in each bin, using an iterative procedure.  Candidate variables were
identified as having unusually large ($>3\sigma$) rms values in all three
of the filters.

\section{Results}

\subsection{Detected variables}

Thirteen candidate variables were discovered using the above method.  All
but three of these stars have CL $>$ 2.0 in the F110W image.  The second
most isolated candidate variable (at 194.05, 197.78 in orbit \#1
coordinates) has CL=0.37 and is clearly a good candidate variable. This
star will be referred to as V1 and is located at $\alpha (2000) =
17^{\mbox{h}}48^{\mbox{m}}05^{\mbox{s}}.14, \delta(2000) =
-24^{\circ}46'38''.2$. Figure \ref{fig.allmagrms} shows V1 along with stars
lying within 12 pixels of V1 (`a'-`e'). Star `a' is a faint star lying only
2.1 pixels from V1. The overlapping PSF of the variable imposes an apparent
but false signal for its fainter companion. None of the other stars have
convincing variations. Further discussion of V1 will be given in section
\ref{sect.v1}.

\begin{figure}[hbt]
\vspace*{-2.0cm}
\epsscale{1.3}
\hspace*{-1.1cm}
\plotone{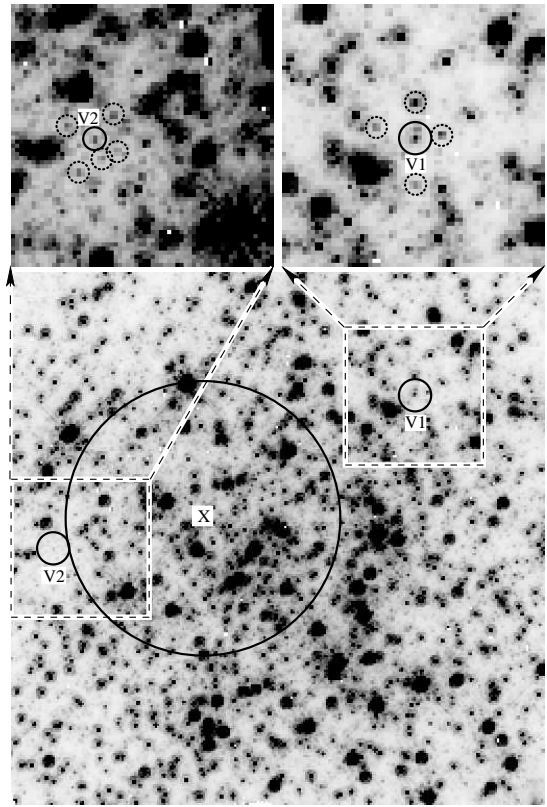}
\vspace*{-1.5cm}
\caption{The F110W image for the first orbit, including the positions of
V1 and V2 and the position and error circle of the X-ray source from
Johnston et al. (1995). The close--up figures include solid circles for the
variables and dotted circles for the near neighbors of V1 (`b'-`e'; `a' is
not circled and is found 2 pixels above V1) and V2.}
\label{fig.fchart}
\end{figure}

A deeper search was also carried out for 2$\sigma$ variations. This
resulted in 90 variable candidates, with 13 stars having CL$ < 2.0$, but
this deeper study revealed only one new, believable variable, henceforth V2
(see section \ref{sect.bluevar}), located at $\alpha (2000) =
17^{\mbox{h}}48^{\mbox{m}}05^{\mbox{s}}.06, \delta(2000) =
-24^{\circ}46'52''.5$. A finding chart (Figure \ref{fig.fchart}) shows the
F110W NIC2 image from the first orbit and the positions of V1 and V2. The
summed image from the first orbit is shown (rather than the dithered,
oversampled image) to show the level of crowding in the single orbit
images.

To test the completeness and depth of our variable search, we carried out
time series simulations of our F110W data, the least crowded images. We
calculated aperture photometry time series for a grid of positions and
added sinusoids with an amplitude of 0.3 mag to these time series. We
considered a fake variable to be ``detected'' if the rms was more than
twice the rms without adding in the sinusoid. This technique automatically
takes crowding into account.  We then tested detection efficiency as a
function of radial distance and absolute magnitude of the fake variables
(for periods between 0.5 hours and 15 hours, the detection efficiency is
constant to within a few percent). The $M_V$ calibration was based on the
assumption that the simulated stars belong to the main sequence or the
giant branch, using stellar isochrones (see Section \ref{sect.v1phot}).
The percentage of detected 0.3 mag simulations as a function of radial
distance from the center of the cluster is shown in Figure
\ref{fig.simul1}. Note the dramatic improvement in detection efficiency as
one moves away from the center of the cluster, but the lack of sensitivity
to variables with $M_V > 4$ (just above the main sequence turn-off; MSTO).

Variable searches were also attempted using analysis of the Lomb-Scargle
periodogram (Scargle 1982) of each time series.  However, the 8 orbits were
spread out over $\sim$1.3 days, and the noise from orbit to orbit is larger
than the noise within a single orbit, because of the dithering. Therefore,
many false signals with periods of a few hours were found in the power
spectra and this method proved to be less efficient at finding candidate
variables than working in the time domain.

\subsection{V1: RR Lyrae Variable}
\label{sect.v1}

We applied several extra tests to V1 to confirm its variability. Because a
bad pixel at (197,203) falls within the V1 aperture for the 4th and 5th
dither positions (1.2 and 1.5 pixels away respectively), we verified that
there is no clear drop in the flux for these two dither positions. Besides
these two dither positions the average offset of the bad pixel from V1 is
3.9 pixels, well outside our aperture.  As a second test, we placed a grid
around V1 and calculated time series for apertures centered on this grid.
Variability was only found for apertures within a few pixels of V1.
Finally, we verified that V1's light curve from PSF-fitting was similar to
that for aperture photometry, and that the quality of the fits was
reasonable compared to other stars.

Henceforth we consider V1 to be a true variable. We now discuss the
photometry and time series for this star.

\subsubsection{Photometry}
\label{sect.v1phot}

Figure \ref{fig.hpcmd} shows the F110W vs F110W-F187W CMD for the central
NIC2 field, from Cohn et al. (2000). The labels represent the variable V1
and its near neighbors (`a'-`e'). The variable is clearly a blue star,
while the stars labeled with `*' are other possible blue stars, many of
which are likely to be red giants scattered into the blue region of the CMD
by errors. The solid line shows the stellar isochrones of Bergbusch and
Vandenberg (1992).  We extrapolated the highest metallicity stellar
isochrones ([Fe/H] = --0.47) to solar metallicity, fed the $T_{eff}$ and
log g values into Kurucz atmospheres and used SYNPHOT to estimate the
NICMOS magnitudes of these Kurucz models, using the reddening measured by
Cohn et al. (2000). Besides the main sequence and giant branch we also show
simple models of the horizontal branch (HB), assuming a large spread in
effective temperature. Since the variable lies close to an extension of the
HB, a RR Lyrae source for the variations is suggested. We also show a model
BS sequence (guided by $T_{eff}$ and log g values from Shara, Saffer and
Livio (1997) for a bright BS in 47 Tuc). Our BS sequence extends to
$\sim$2.2 magnitudes above the MSTO in an unreddened $V$ vs $B-V$ CMD,
about equal to the extent of the BS sequence in the center of 47 Tuc
obtained with {\it HST} (Gilliland et al. 1998).

Despite the very small FOV of the NIC2 camera there is some field star
contamination because the line of sight towards Terzan 5 is close to the
galactic center ($L$=3.81, $B$=1.67). We estimated the field star component
using the ground-based observations of Ortolani et al. (1996). Using their
SUSI ($2.2'\times2.2'$ FOV) observations extending to about 1.5 magnitudes
below the HB (see their Figure 5) we defined field stars to have $V-I <
3.2$.  Scaling to the NIC2 FOV we predict 4--5 bright field stars should be
found in our field. This is in excellent agreement with Figure
\ref{fig.hpcmd} showing 5--6 likely field stars with F110W $<$ 19.0 and
F110W-F187W $<$--0.8.

\subsubsection{Time series}

Figure \ref{fig.ts1-6} shows the aperture photometry light curves of V1
along with light curves of stars `a'--`e'.  The F187W and F187N light
curves have been shifted vertically so that the apparent peaks of the light
curves (in orbit 3) match up.  Besides the overlapping star `a', V1 clearly
stands out from its near neighbors.  A distinctive element of the light
curve of V1 is its asymmetry, typical behavior for RR Lyraes of type
ab. Motivated by this, we carried out fits of our NICMOS light curves to
two sets of IR light curves for RR Lyraes, $K$-band templates from Jones,
Carney and Fulbright (1996) and the $JHK$ light curves of Cacciari,
Clementini and Fernley (1992) for three field RRab Lyraes. We varied the
period, magnitude zero-point and amplitude of the IR light curves (tabulated
as functions of phase) to minimize the residual between our light curves
and the tabulated light curves. Since the F187W and F187N filters lie
roughly halfway between the ground-based $H$ and $K$ passbands we used both
$H$ and $K$ data to fit to our combined long-wavelength (F187W and F187N)
light curves. We used $J$ data to fit to our F110W light curves.

Jones et al. (1996) give templates for (a) RRc variables, (b) RRab
variables, $A(B) < 1.0 $mag, (c) RRabs, $1.2 < A(B) < 1.3$, (d) RRabs,
$A(B) \approx 1.5$, and (e) RRabs, $1.6 < A(B) < 1.7$. By examining the
residuals and comparing them with the expected scatter we can reject (a) at
the $\sim 3.3 \sigma$ level (the derived amplitude of 0.21 mag is also too
high by about a factor of 2). The best fits were obtained for the (c), (d)
and (e) templates, with little difference between these three. With (d) for
example, we derived a period of 0.60 days and a peak-to-peak amplitude of
0.25 mag when fitting to the 1.87$\mu$m filters.

The Cacciari et al. (1992) light curves for W Tuc gave the best fit to our
data. For $J$ we derived a period of 0.63 days and an amplitude of 0.44 mag
when fitting to F110W. Using $H$ and $K$ time series when fitting to the
1.87$\mu$m filters, we derived periods (amplitudes) of 0.61 days (0.24 mag)
and 0.61 days (0.25 mag) respectively. The ratio between the amplitudes
derived for $J$ and for $K$ was only 15\% larger than the ratio for the
field systems, a reasonable agreement given the differences between the
{\it HST} and ground-based filters.  The similarity between the derived
periods shows that the error for the period determination is likely to be
only 5\%-10\%. The rms of the residuals from these fits are 0.018 (Jones
$K$), 0.021 (Cacciari $J$), 0.017 (Cacciari $H$) and 0.016 (Cacciari
$K$), consistent with the noise properties of stars in a similar magnitude
range (see Figure \ref{fig.allmagrms}).

Figure \ref{fig.lc} shows the results of these fits.  Note the similarity
between the $H$ and $K$-band template light curves, the close
correspondence between the F187W and F187N data and the rapid rise time for
the F187N data near 7000s (matching well with the template light curves).

\subsection{A Search for Blue variables: V2}
\label{sect.bluevar}

Along with the general search for variables in the NIC2 images, we also
paid special attention to stars with blue colors, in an attempt to find the
infrared counterpart to the LMXB known in Terzan 5, and to search for other
RR Lyrae variables and possible bright BS variables.

To select a robust sample of stars for time series analysis, only the stars
with CL $ <1.5$ in the F187W image were examined in greater detail. To
compare our photometry with that of Cohn et al. (2000), we show in Figure
\ref{fig.mycmd}(a) the F110W vs F110W-F187W CMD and in Figure
\ref{fig.mycmd}(b) the F187W vs F187N-F187W CMD based on the flux values
derived from our time series.  The labeled stars from Figure
\ref{fig.hpcmd} are shown, plus extra stars with apparently blue colors
from our photometry (labeled with `+'). Note that: (i) the RR Lyrae
variable is found in a similar part of the F110W vs F110W-F187W CMD as
Figure \ref{fig.hpcmd}, (ii) many of the blue stars labeled `*' from
Figure \ref{fig.hpcmd} are missing from Figure \ref{fig.mycmd} because they
are too contaminated, and (iii) of the blue stars that do appear, most of
them, reassuringly, have blue colors. Apparent P$\alpha$-bright objects in
Figure \ref{fig.mycmd}(b) are caused by errors from near neighbors.

Figure \ref{fig.rms-confil} shows the corresponding plots of rms versus
magnitude. Using an uncontaminated sample of stars causes a significant
reduction in the number of false variables appearing in these plots. In
particular the RR Lyrae variable and its near neighbor stand out much more
clearly from the other stars. The star labeled with `V2' also has an
unusually large rms, especially in the F110W filter.  This large rms is
unlikely to be caused by near-neighbor effects, since this star is well
isolated from other stars, having CL = 0.05, 0.24 and 0.22 in F110W, F187W
and F187N respectively. (The dithered images show a faint star about 2
pixels away from V2 that is too faint to be included in the master star
list. However, this star makes $<$ 10\% contribution to the counts in V2's
aperture.)  Also, there are no bad pixels or other obvious artifacts near
this star. Figure \ref{fig.extrav} shows plots of the time series for V2.

The absolute magnitude of V2 is uncertain because we lack optical
information for this star and only have limited color information. Noting
that V2 lies near the end of our BS sequence in Figure \ref{fig.hpcmd}, we
assume that a 8000 K Kurucz atmosphere gives a reasonable approximation to
the spectrum of V2, and thus estimate that $M_V \sim 2.0$.

Power spectrum analysis gave periods and amplitudes of (6.94 hr, 0.27 mag;
F110W), (6.93 hr, 0.33 mag; F187W) and (6.88 hr, 0.27 mag; F187N), assuming
the variation is sinusoidal.  By adapting the fitting technique used for
the RR Lyrae variable (using a sinusoid for the template), we estimated the
periods and amplitudes to be (6.96 $\pm$ 0.20) hr and (0.27 $\pm$ 0.09) mag
(F110W), (6.74 $\pm$ 0.25) hr and (0.40 $\pm$ 0.19) mag (F187W) and (6.85
$\pm$ 0.6) hr and (0.29 $\pm$ 0.27) mag (F187N) respectively, in good
agreement with the power spectrum results quoted above.  Using this fitting
procedure, the rms of the residuals were found to be 0.045 mag, 0.062 mag
and 0.122 mag for the three filters. These values agree well with the mean
rms values (0.045 $\pm$ 0.010, 0.069 $\pm$ 0.026 and 0.111 $\pm$ 0.032)
determined for the bins containing V2 in Figure \ref{fig.rms-confil}. While
a slightly larger amplitude is suggested for F187W, the errors for this
filter are quite large, and a single sinusoidal fit to all of the data is
allowed by the errors. For example, adopting a `compromise' solution for
all 3 filters of period 6.98 hours and amplitude 0.282 mag gives rms
residuals of only 0.049, 0.075 and 0.126 mag, comfortably within the
1$\sigma$ limits quoted above.

\subsection{Millisecond pulsar}

\subsubsection{Astrometry}

The position of the MSP in Terzan 5 is known to within 0.08$''$ (Fruchter
\& Goss 2000) in the radio frame of reference (which uses quasars), but
positional information given in the NICMOS data files is accurate to only
about 1$''$ in absolute terms, and is limited by uncertainties in the
locations of the instrument apertures relative to the Optical Telescope
Assembly axis and the inherent uncertainty in the positions of the two
guide stars used. For {\it HST} images, differential astrometry (measuring
relative positions within an image) is much more accurate. We therefore
carried out an astrometric study of the Terzan 5 field to improve the
absolute accuracy of the positions, and cut down the size of the error
circle for the MSP. Our positions should then only be limited by
uncertainties in our astrometric solution and by systematic errors in the
STScI Digitized Sky Survey (DSS).

We used two steps to carry out this astrometric study: (i) We matched up
the maximum possible number of stars (9) between the DSS and the study of
Ortolani et al. (1996), and used a linear regression in x and y to
calculate an astrometric solution for the Ortolani stars. (ii) We then
performed a plate solution match between the Ortolani and NICMOS fields. A
direct solution between the DSS image and the NICMOS fields was not
possible because the small FOV of NICMOS and poor resolution of the DSS
images limited the number of matching stars.

The position of the MSP is:

\noindent $\alpha (2000) = 17^{\mbox{h}}48^{\mbox{m}}02^{\mbox{s}}.21,
\delta(2000) = -24^{\circ}46'37''.2$

\noindent (Fruchter \& Goss 2000).  Using the two steps
described above, the nominal MSP position corresponds to pixel coordinates
of (141.6, 180.5) for orbits 1--4, only 0.12$''$ from the predicted
position using the header information. The rms residuals after applying the
fit described in step (i) above are 0.341$''$ and 0.140$''$ in the x and y
directions respectively. The errors for step (ii) are much smaller.

\subsubsection{Photometry}

Nice and Thorsett (1992) estimate that if the MSP companion is a main sequence
star it is likely that 0.09 $M_{\odot} \lesssim M_2 \lesssim 0.11
M_{\odot}$, right near the lower limiting mass for hydrogen burning.
However, by analogy with the eclipsing MSP 1957+20, the optical counterpart
is expected to be much brighter ($M_V \sim 10-11$) than a normal main
sequence star due to reprocessing of pulsar spin-down energy on the
evaporating wind.  A search for photometric variability from the MSP is
nevertheless very challenging given the considerable distance of the Terzan
5 MSP and its high reddening.  These problems were compounded by 3 extra
problems: (1) we had initially planned to observe the cluster center and
the MSP simultaneously using the NIC1 and NIC2 cameras (respectively) in
parallel, given the virtually identical value for the cluster center--MSP
offset and the NIC1--NIC2 offset. However, in phase 2 planning we found
that we were unable to use the required telescope roll angle, and therefore
decided to slew between the two targets, concentrating most of our time on
cluster center and cutting our observing time on the MSP by a factor of
$\sim$4. This strategy effectively reduced the observed amplitude of any
MSP variations that may be present.  (2) The nominal position of the MSP
lies close to a HB star and two bright giants (at relative distances of
0.55$''$, 1.3$''$ and 2.5$''$; see Figure \ref{fig.mspfchart}). The extra
background from the PSF halos of these bright stars compromised the MSP
observations. (3) The NIC1 data is much more affected by transient
artifacts, including `persistence', than the central NIC2 data.

Despite these limitations the MSP--field data is less crowded than the
central NIC2 data, enabling variability studies of stars at and below the
MSTO. Our photometric analysis was similar to that carried out for the
central NIC2 field except that we did not have the advantage of
cross-checking candidate variables between different filters.  A plot of
time series rms versus magnitude for the MSP field is shown in Figure
\ref{fig.rmsmsp}. The HB stars identified from the $V$ vs $V-I$ CMD of
Ortolani et al. (1996) are labeled, as are stars (and in some cases PSF
tendrils) lying within 1$''$ of the nominal MSP position, and the two
giants near the position of the MSP. Note that the stars near the MSP have
slightly higher noise levels than average because of the higher background
near the bright giant stars.

We carefully examined the results to search for evidence of variability,
but no authentic variable candidates were found.  We also carried out a
second dedicated search for variability from the MSP by searching a grid of
positions within a 3$\sigma$ (1.1$''$) radius circle around the measured
position of the MSP. Two criteria were used in this variability search:
excess rms in a small group of adjacent pixels and high linear correlation
of the 8 point time series with the predicted time series of the MSP
(assuming a sinusoidal shape for the light curve and using the ephemeris of
Nice and Thorsett 1992). No candidates for a variable MSP were found.

To estimate the depth of our MSP variability search, we used our Kurucz
atmosphere/synphot simulations described above. We simulated MSP variations
using sinusoidal light curves (at the MSP period of 1.82 hr) with an
amplitude of 1 mag and added these light curves to our measured light
curves for the above grid. A MSP simulation was considered `detected' if
the linear correlation between the predicted and simulated light curves was
$>$0.5 and the rms of the simulated light curve was 50\% higher than the
original rms. At $M_V=3.9$ (F140W=21.3) only 43\% of the simulations were
detected and at $M_V=3.4$ (F140W=20.0) 86\% of the simulations were
detected.  To show the effect of the high sky levels near the expected
position of the MSP we carried out another set of simulations for a grid
centered on a point well away from the bright giants.  Here, our
variability search was able to go 1.5 mag deeper, but still falls well
short of the desired levels to detect variability, assuming that the
secondary has $M_V \sim 10-11$.

\section{Discussion}

\subsection{RR Lyrae variable}

The identification of V1 as a RR Lyrae variable and a cluster member is
secure, based on the light curve, the position of the star in the CMD and
the small chance of field star contamination.  As with 47 Tuc, only one RR
Lyrae has been discovered in Terzan 5 and the bulk of the HB stars are red
and lie near the giant branch (see Figure \ref{fig.hpcmd}), as they do in
most metal-rich globular clusters. However, Cohn et al. (2000) report the
possible detection of a small number of blue HB stars in Terzan 5. The high
reddening and crowding in this cluster makes such detections much more
difficult than in NGC 6388 and NGC 6441 where many blue HB stars have been
found and there are multiple RR Lyrae variables.  Rich et al. (1997) have
explored two possibilities to explain the blue HB star populations in NGC 6388
and NGC 6441, the effects of age and dynamics. At high metallicity an
increase in age can populate a blue HB tail with stars at the low end of
the mass distribution, while dynamical encounters in the cores of dense
clusters can result in extra mass loss via tidal stripping. Rich et
al. (1997) conclude that although neither effect alone appears sufficient
to explain the cluster blue HBs, a combination of them may have an effect.

Our discovery of a RR Lyrae variable in Terzan 5 is therefore useful for
the study of the influence of metallicity, age and interaction rate on the
production of HB stars in the instability strip. For example, Terzan 5 has
a higher metallicity than any of the above clusters which should result in
fewer RR Lyraes, but it also has a higher interaction rate which may result
in a higher number of RR Lyraes (see Rich et al. 1997). It is possible that
these two factors combine. For example, maybe a large number of stars in
Terzan 5 lose mass via tidal stripping but the cluster metallicity is high
enough that only the stars with the most extreme mass loss show up as RR
Lyraes.

Sweigart and Catelan (1998) note that the HBs in NGC 6388 and NGC 6441 have
a pronounced upward slope with decreasing $B-V$, a result which is not
predicted by canonical HB models and which cannot be reproduced by either a
greater cluster age or enhanced mass loss along the red giant branch. They
explore three possibilities to explain these sloped HBs: (1) a high helium
abundance, (2) a rotation scenario in which internal rotation during the
RGB phase increases the HB core mass, and (3) a helium-mixing scenario in
which deep mixing on the RGB enhances the the envelope helium abundance. 

One prediction of the high-Y scenario is that the number ratio $R$ of HB
stars to RGB stars brighter than the mean RR Lyrae luminosity should be
large. We calculated $R$ for Terzan 5 by defining the HB as those stars
having 15.6 $<$ F110W $<$ 16.1 and correcting for the large number of RGB
stars found in this region of the CMD by averaging and subtracting the
number of stars present in adjacent magnitude bins (15.1 $<$ F110W $<$ 15.6
and 16.1 $<$ F110W $<$ 16.6). Our value for $R$ is 1.03 $\pm$ 0.11, where
the error is based on Poisson statistics alone. Although systematics will
increase the error, this value is clearly inconsistent with the high-Y
scenarios presented in Sweigart and Catelan (1998), where $R=3-4$. Layden
et al. (1999) came to a similar conclusion for their study of NGC
6441. Sweigart and Catelan (1998) suggest a variety of other observational
tests for discriminating among the various scenarios, all requiring further
observations.

Figure \ref{fig.hpcmd} shows that V1 may be somewhat brighter ($\sim$ 0.3
mag) than expected for a blue HB star, consistent with the sloped HB slopes
mentioned above, although uncertainties in the large reddening for this
cluster and the limitations of our simple HB model make detailed
comparisons difficult.  As Sweigart and Catelan (1998) point out, unusually
bright RR Lyraes should have unusually long periods, as observed for the RR
Lyraes in NGC 6388, NGC 6441 and 47 Tuc. This self-consistency between the
RR Lyrae luminosities and periods is reassuring.

The obvious question arises: is the period of V1 unusually long?  First,
the metallicity has to be considered. V1 is likely to be the most
metal-rich RR Lyrae yet found in a globular cluster, since Terzan 5 is
thought to have roughly solar metallicity (Ortolani et al. 1996) or even
higher (Cohn et al. 2000 estimate [Fe/H]=0.25). NGC 6388, NGC 6441 and 47
Tuc have [Fe/H] values of --0.60, --0.53 and --0.76 respectively (Harris
1999)\footnote{The Catalog of Parameters for Milky Way Globular Clusters,
compiled by William E. Harris, McMaster University, revised June 22, 1999
is available at http://physun.mcmaster.ca/$\sim$harris/mwgc.dat}. It may
even be the highest metallicity RR Lyrae ever found.

Examination of the metallicity--period relationship for field RR
Lyraes (Layden 1995) shows that V1 has a significantly longer period than
field stars with similar metallicity. The period-metallicity distribution
of Pritzl et al. (2000) shows that V1 lies well outside the two Oosterhoff
classes, with a longer than expected period given the metallicity of Terzan
5.  Therefore V1 joins the rapidly growing subclass (suggested by Layden et
al. 1999) of long period RR Lyraes in metal rich globular clusters.

Also plotted on Figure 1 of Pritzl et al. (2000) are the periods and
metallicities of the RR Lyraes in NGC 6388 and NGC 6441. These two clusters
are clearly separate from the two Oosterhoff classes, with higher
metallicity and longer periods. Interestingly, while the average period for
these two clusters is longer than {\it all} of the OoI and OoII globulars,
V1 has a period lying almost exactly between the 2 Oosterhoff
groups. Further, a line drawn between V1 and the average of NGC 6388 and
NGC 6441 in the period--metallicity plot lies almost parallel to the line
between the 2 Oosterhoff classes, suggesting two period--metallicity
relations with similar slope may exist.

This interesting but speculative result clearly requires verification by
observing a larger sample of RR Lyraes in clusters with solar metallicity
or above. It could be that V1 has an unusually short period compared to
other (undetected) RR Lyraes in this cluster. For example, the periods in
NGC 6441 alone range from $\sim$0.6 days to $\sim$ 0.9 days, with a
reasonably flat distribution between these two extremes.

\subsection{The faint variable V2: LMXB or eclipsing BS?}

To summarize our results, V2 is blue, relatively faint and variable, with a
near sinusoidal light curve, and a period of $\sim$7 hr. We consider two
possibilities for this variable, that it is a LMXB or that it is an
eclipsing BS.

First, we consider the LMXB hypothesis. In this case the variability could
be caused by at least three different phenomena: (1) Ellipsoidal variations
from distortion of the secondary. These variations typically have a
distinctive double-lobed pattern, with a difference in depth between
successive minima of $\sim$0.05--0.10 mag and an expected difference in
amplitude between F110W and F187W of $\sim$0.05 mag, using reasonable
assumptions regarding mass ratios, temperatures and inclinations (van
Hamme, private communication). These amplitude differences are not clearly
detected in the data, though the relatively large error for the F187W
amplitude means we cannot rule out this possibility.

(2) Another possible explanation of variability, under the LMXB hypothesis,
is the effects of heating of the secondary by the disk and neutron
star. Here, the expected light curve is also double-lobed, but with the
positions of the deep and shallow minima reversed (see van Paradijs \&
McClintock 1995). A large contribution from heating would be obvious in the
light curve because of the significant expected difference in eclipse
depth, but a small contribution would plausibly cancel out the difference
between the depths of adjacent minima, when combined with an ellipsoidal
variation. However, this requires some fine-tuning of the binary parameters
and a difference between the light curve amplitudes remains (Van Hamme,
private communication). Note that if either case (1) or (2) applies to V2
then the period would be twice the 7-hour sinusoidal variability period, ie
$\sim$14 hours. It would thus be similar to the field LMXB, and recurrent
transient, Aql X-1 (Chevalier \& Ilovaisky 1991), an attractive possibility
given the transient, or highly variable nature of the Terzan 5 LMXB.

(3) Flickering from the disk may also be expected to show up in the light
curve of a LMXB, as it does in field systems. If V2 is a LMXB, the blue
color of V2 would imply a reasonably prominent disk (the source of
flickering).  However, no evidence for flickering is seen as the residuals
after subtracting the sinusoidal model are consistent with the noise
properties of the data.

Besides being blue, variable objects, LMXBs also typically have weak
emission lines in the optical and infrared.  Shahbaz et al. (1996) show
that the average H$\beta$ equivalent width (EW) of 15 LMXBs, is 1.9
\AA. Scaling by the average EW ratio between H$\beta$ and P$\alpha$ for the
two CVs PQ Gem and EX Hya (using values from Dhillon et al. 1997; no
published P$\alpha$ EWs for LMXBs were found) we expect an average
P$\alpha$ EW of 3.8 \AA. This is equivalent to F187N-F187W = --0.07, well
below the 1$\sigma$ error for the faint end of our F187N vs F187N-F187W CMD
(containing V2) of 0.11 mag. Therefore the P$\alpha$ data is inconclusive
in determining whether V2 is an LMXB.

An obvious constraint on the identify of V2 is its position relative to the
X-ray source identified by Johnston et al. (1995). Our measured positional
offset of 5.5$''$ is just outside the 5$''$ error circle quoted by Johnston
et al. (1995). We hope to significantly reduce the size of this error
circle with the use of our Cycle 1 Chandra observations. With these deep,
50ks observations the LMXB should have an error circle $< 1''$
and should be distinguishable from the large expected population of CVs and
MSPs in this cluster, unless it is in deep quiescence.

\begin{figure}[hbt]
\vspace*{-2.0cm}
\epsscale{1.0}
\plotone{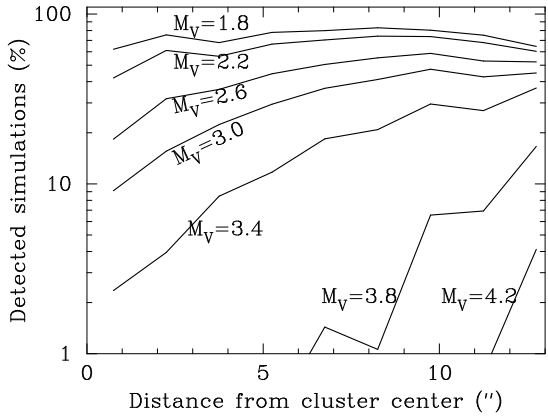}
\caption{The percentage of detected simulations (with 0.3 mag amplitude) as a
function of radial distance from the center of the cluster. A range of
absolute magnitudes were used for the simulations. The drop in detection
rate beyond a radial distance of $\sim 10 ''$ is because of the decreased
fraction of annulus that lies in the NIC2 field.}
\label{fig.simul1}
\end{figure}

We may also compare the period and luminosity of V2 with UV or optically
identified globular cluster LMXBs.  Three of the six globular cluster LMXBs
with optical counterparts have $M_V \gtrsim$4.5, according to Deutsch,
Margon and Anderson (2000).  Figure \ref{fig.simul1} shows that this is
beyond our detection limits, for stars on the main sequence or the giant
branch (for hotter objects our detection limits are even shallower). To
make matters worse the Terzan 5 LMXB, being a transient should be fainter
than most other cluster LMXBs.  We can predict $M_V$ using the dependence
of optical luminosity upon X-ray luminosity and period (van Paradijs and
McClintock 1994) a relation that holds for both cluster and field systems
as shown by Deutsch et al. (2000). For a period of $\sim$14 hrs, we
estimate $M_V=4.2$, beyond our detection limits. Finally, we note that 4 of
the 5 systems with optical counterparts and measured periods have period
$\lesssim$ 5.7 hours and three have periods $\lesssim$ 0.73 hours.

\begin{figure}[hbt]
\epsscale{1.0}
\plotone{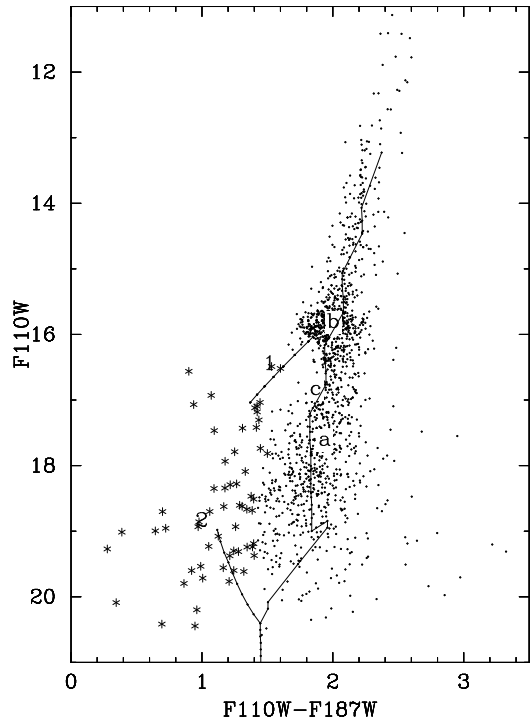}
\caption{F110W vs F110W-F187W CMD from Cohn et al. (2000). The numbered and
labeled stars discussed in the text are shown, along with a selection of
blue stars (`*').  Isochrones from Bergbusch and Vandenberg (1992) are
shown by the solid line, plus HB and BS sequences. The jagged appearance of
the isochrones is because they have been fed through Kurucz atmospheres,
quantified in $T_{eff}$ and log g, to calculate the NICMOS magnitudes.}
\label{fig.hpcmd}
\end{figure}

\begin{figure}[hbt]
\epsscale{1.0}
\plotone{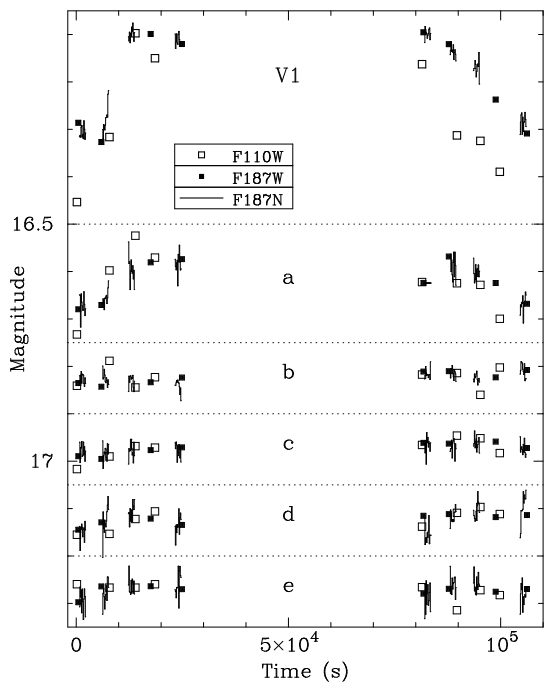}
\caption{Time series for V1 plus 5 near neighbors. The unfilled squares
show the F110W data (averaged over each orbit), the filled squares the
F187W data (orbital average) and the solid lines the F187N data. For V1 the
zero-points have been adjusted so that the F187W and F187N filters have the
same magnitude for orbit 3 as F110W. For the neighboring stars the
zero-points have been adjusted so that the average magnitude of the three
filters over all of the orbits are equal. A further correction was applied
to offset the time series from each other.}
\label{fig.ts1-6}
\end{figure}

\begin{figure}[hbt]
\vspace*{-2cm}
\epsscale{1.0}
\plotone{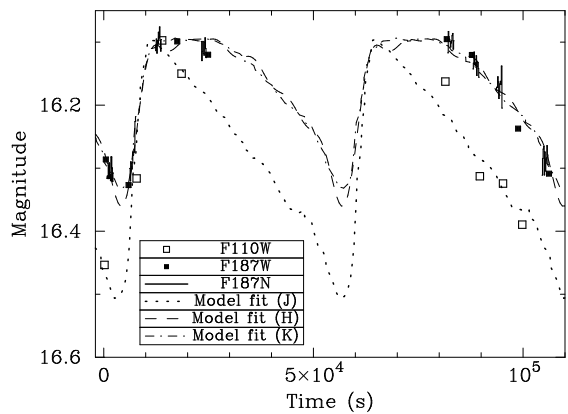}
\caption{Time series for the RR Lyrae variable (V1). The unfilled squares
show the F110W data, the filled squares the F187W data and the solid lines
the F187N data.  The fitted $J$, $H$ and $K$ templates are shown by the
dotted, dashed and dot-dashed lines respectively..}
\label{fig.lc}
\end{figure}

\begin{figure}[hbt]
\vspace*{-2cm}
\plotone{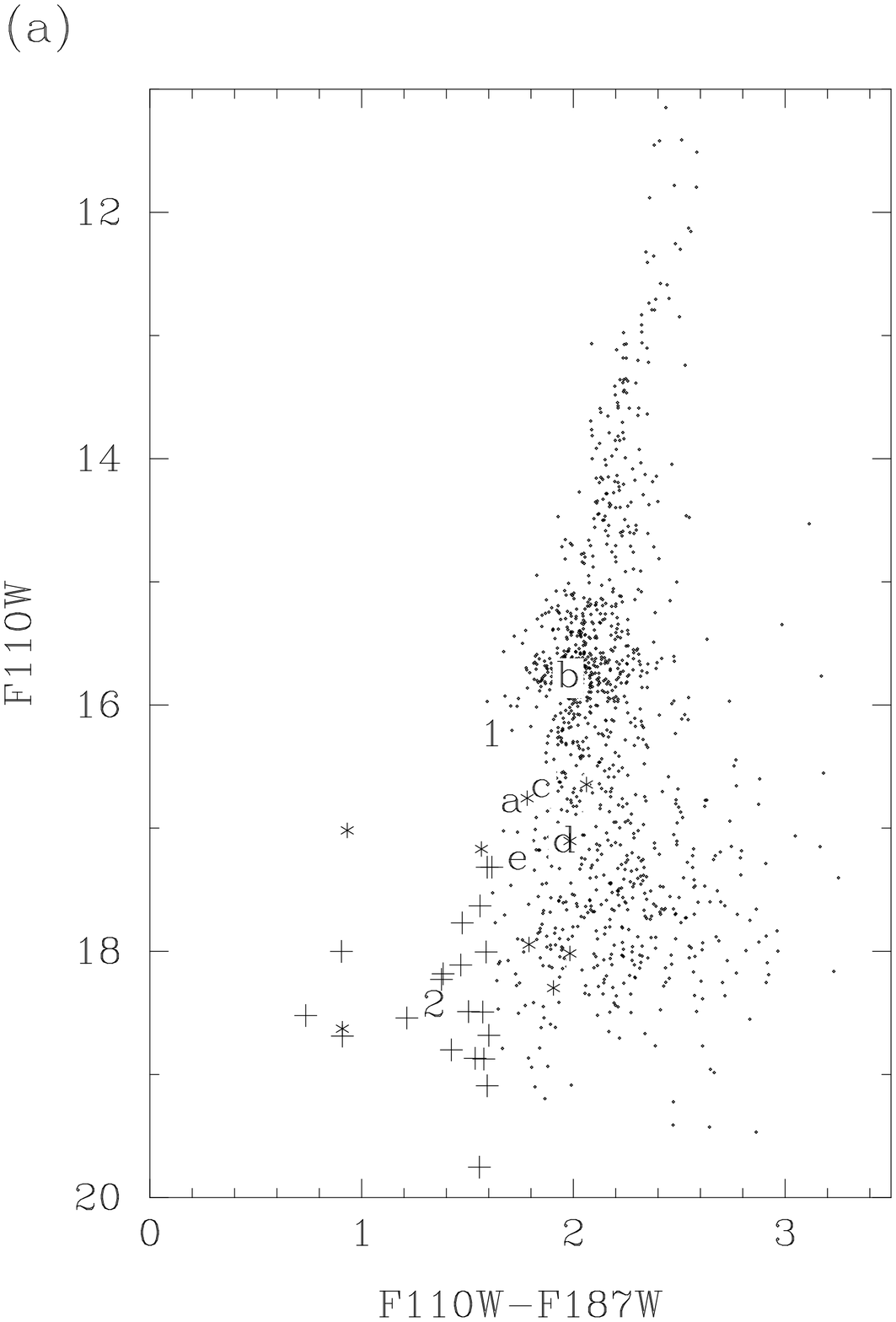}
\plotone{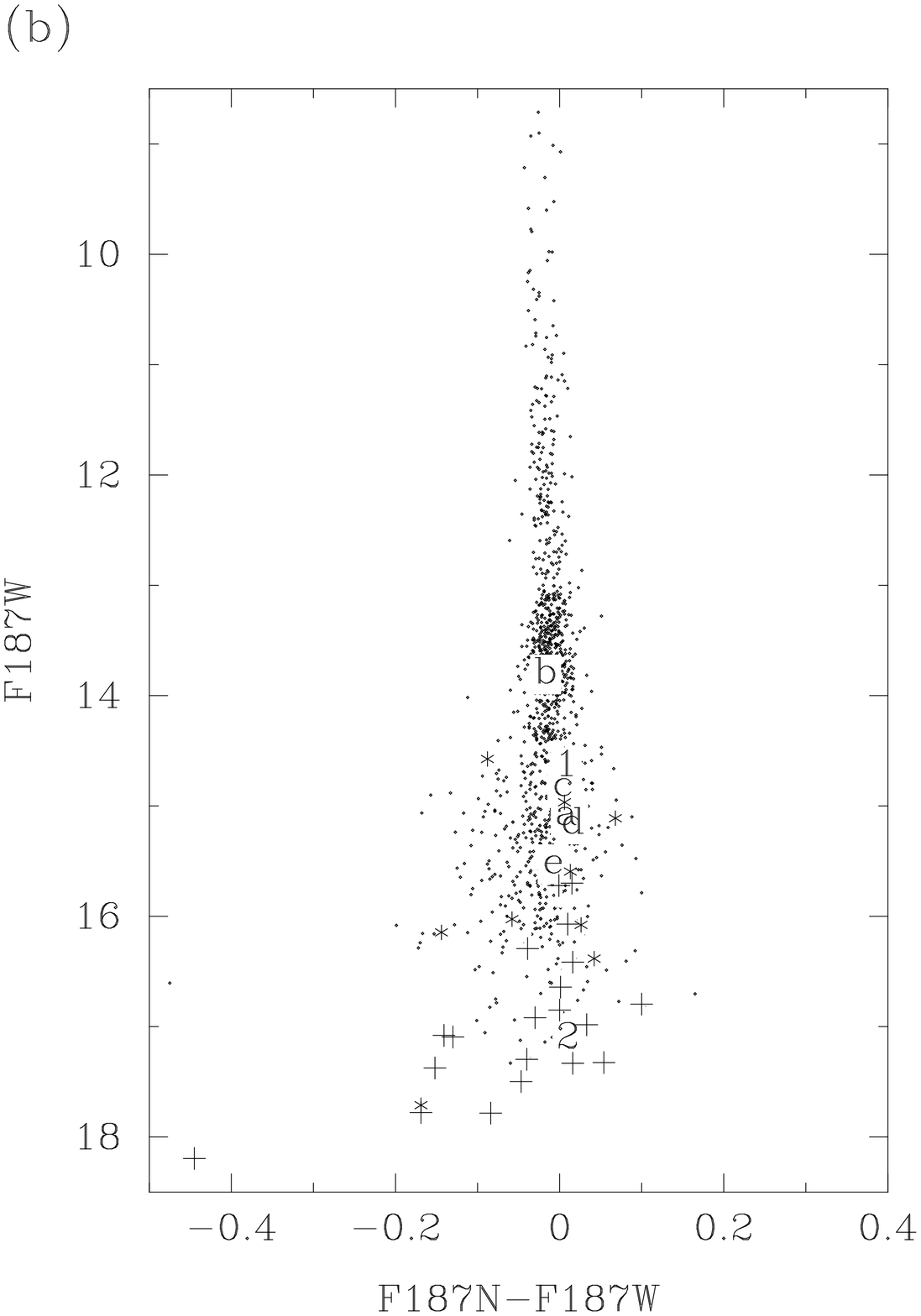}
\vspace*{-1cm}
\caption{CMDs from combining the time series photometry.  Figure
\ref{fig.mycmd}(a) shows the F110W vs F110W-F187W CMD and Figure
\ref{fig.mycmd}(b) shows the F187W vs F187N-F187W CMD.  Only stars with CL
$ < 1.5$ in F187W have been plotted here. The numbered and labeled stars
from Figure \ref{fig.hpcmd} are shown along with extra blue stars from our
photometry (`+').}
\label{fig.mycmd}
\end{figure}

\begin{figure}[hbt]
\epsscale{1.0}
\plotone{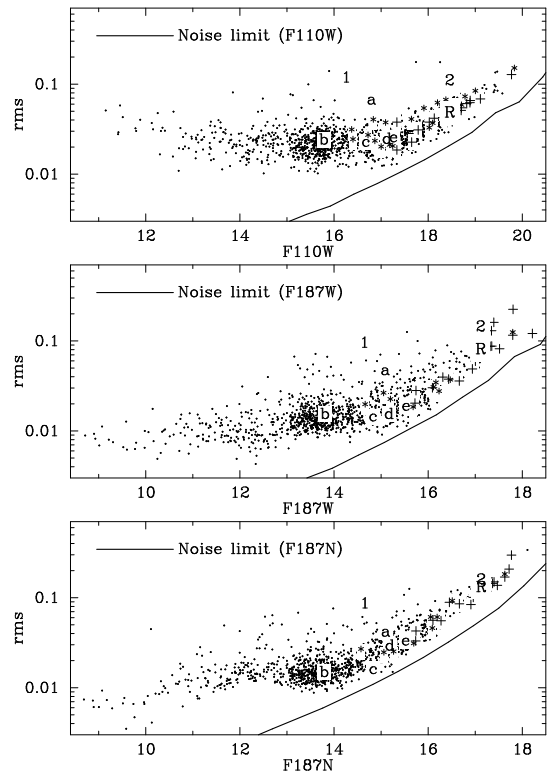}
\caption{A similar plot to Figure \ref{fig.allmagrms} except that only
stars with CL (F187W) $<$ 1.5 have been plotted. The variable V2 has been
added to this figure, along with a symbol `R' showing the residual after
subtracting a sinusoidal fit to the V2 light curve.}
\label{fig.rms-confil}
\end{figure}

\begin{figure}[hbt]
\epsscale{1.0}
\plotone{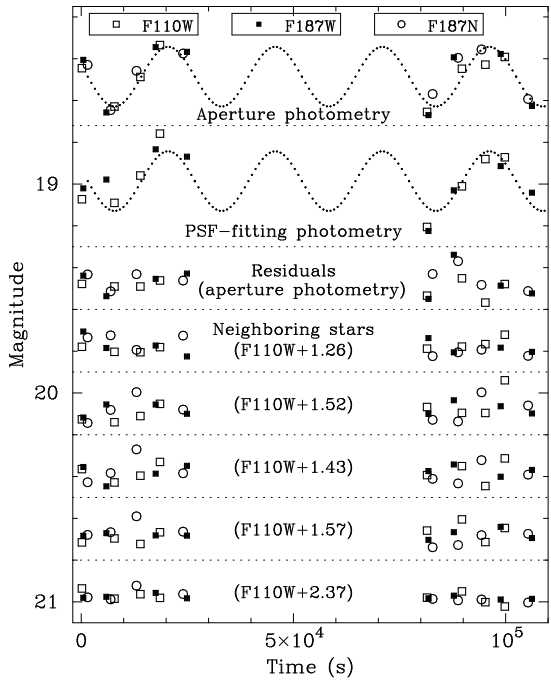}
\caption{Light curves for V2. The upper panel shows the aperture photometry
results (including a sinusoidal model fit to the data), the next panel the
noiser PSF-fitting results, then the residuals after subtracting the
model. The other panels shows the light curves of 5 nearby stars, with the
required F110W offset shown in parentheses.}
\label{fig.extrav}
\end{figure}

\begin{figure}[hbt]
\vspace*{-5cm}
\epsscale{1.3}
\hspace*{-1.1cm}
\plotone{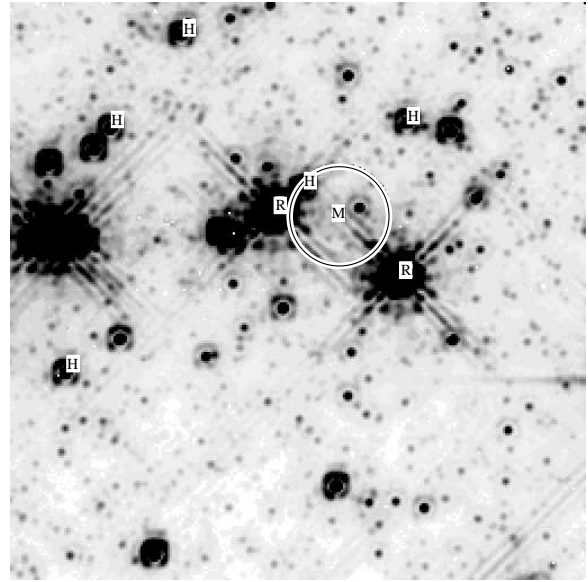}
\vspace*{-3cm}
\caption{A finding chart for the NIC1 MSP field. The nominal MSP position
is shown (`M'), along with HBs (`H') and red giant stars (`R') discussed in
the text. A 1$''$ circle is plotted around the nominal MSP position.}
\label{fig.mspfchart}
\end{figure}

\begin{figure}[hbt]
\vspace*{-5cm}
\epsscale{0.8}
\hspace*{-1cm}
\plotone{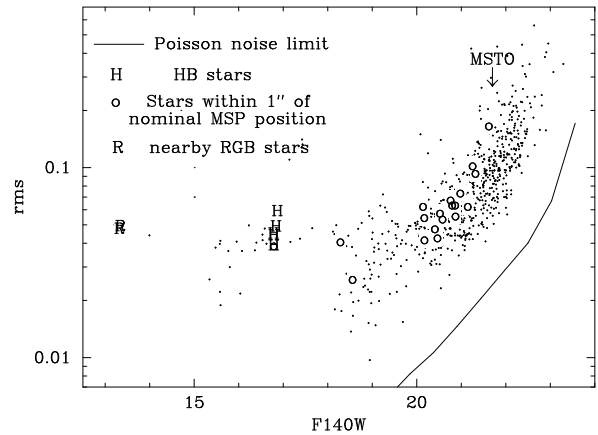}
\vspace*{-0cm}
\caption{The plot of rms versus magnitude for the NIC1 field containing the
MSP. Stars within 1$''$ of the nominal MSP position are shown, along with
HB stars and red giant stars discussed in the text. The estimated position
of the turnoff is shown.}
\label{fig.rmsmsp}
\end{figure}

To summarize, although we find no {\it individual} argument gives
compelling evidence against V2 being the infrared counterpart of the Terzan
5 LMXB, the accumulated weight of evidence argues against it. An
alternative hypothesis is that V2 is an eclipsing BS. This is consistent
with: (1) The position of V2 in the F110W vs F110W-F187W CMD (see figure
\ref{fig.hpcmd}). V2 has about the same luminosity as the brightest BS in
47 Tuc, and its luminosity may also be significantly enhanced by its binary
companion. (2) The light curve amplitudes of V2. Sinusoidal light curves
with our measured amplitudes are easy to reproduce using the light curve
models of Wilson for W UMa systems (van Hamme, private communication). A
growing number of binary BSs are being discovered in globular clusters (eg
Edmonds et al. 1996 and Kaluzny et al. 1998), although the fraction of BSs
that are binaries is typically quite low.  (3) The inferred orbital period
of 14 hours is consistent with the periods of other binary BSs in globular
clusters such as 47 Tuc (Edmonds et al. 1996). As with the LMXB hypothesis,
the P$\alpha$ data is inconclusive, since for a model BS with $T_{eff}$ =
8000 K (at the top of our artificial BS sequence), F187N-F187W = 0.04, well
within our 0.11 mag scatter for this line discriminant.

Using our simulations we can ask how many eclipsing binaries we should
expect to see with our data set.  Since most eclipsing binaries discovered
in clusters are found near the MSTO (eg Edmonds et al. 1996) and we have
very little sensitivity to such variables, the simple answer is none.
Therefore, a naive calculation may predict a very large number of eclipsing
systems (and BSs) should be found in Terzan 5, if V2 is indeed an eclipsing
BS. While clearly small number statistics apply since we have detected only
one system, it is striking that evidence exists for a bright eclipsing BS
at the edge of our detection limits. A relatively large number of BSs and
binaries would have interesting consequences for the dynamics of this
extreme globular cluster. For example, a high BS number may be caused by a
high stellar interaction rate, which itself may be enhanced by a large
binary fraction. A high interaction rate is expected for Terzan 5 because
of its moderately sized core and extremely high stellar density (Cohn et al.
2000). Terzan 5 also may be the globular cluster with the largest
population of MSPs in the galaxy (Fruchter \& Goss 2000), further evidence
for possible large interaction rates.

It is useful to discuss the potential for detailed studies of other
interaction products or binaries in Terzan 5. Our upcoming GO Chandra
observations of Terzan 5 should allow a deep search for CVs in this
cluster, unless the LMXB is in outburst. Unfortunately optical or IR
identifications of such sources will be extremely difficult or
impossible. We have already seen the problems with IR observations, where
crowding prevented us from reaching even close to the MSTO, but with their
smaller PSFs, Gemini and NGST will offer substantial improvements over {\it
HST} for studying this cluster.  In optical passbands such as $V$, the PSF
will be substantially reduced in size but the enormous reddening for this
cluster implies that the MSTO will be found near $V \sim$25.5, and
therefore CVs at $V\gtrsim 28$.  Studies of the BS sequence will be
difficult but may be feasible depending on the results of detailed crowding
simulations.

\acknowledgments

We acknowledge assistance from Mark Dickinson, Torsten Boecker and Howard
Bushouse at STScI in recalibrating the F187N data, and assistance from
Charles Bailyn, Adrienne Cool and Andy Fruchter in planning the
observations. We also thank Walter van Hamme for advice about binary light
curve modeling, and Ron Gilliland for useful discussions. The authors would
like to thank the referee Mike Shara for his helpful comments on the
manuscript. This work was partially supported by HST grant GO-7889.



\newpage

\clearpage


\end{document}